\documentclass[useAMS,usenatbib]{mn2e}
\usepackage{graphicx}

\newcommand{\spitzer}{{\textit Spitzer}}
\newcommand{\aap}{A$\&$A}
\newcommand{\aj}{AJ}
\newcommand{\apj}{ApJ}
\newcommand{\apjl}{ApJ}
\newcommand{\apjs}{ApJS}
\newcommand{\mnras}{MNRAS}

\newcommand{\nat}{Nature}
\newcommand{\pasp}{PASP}

\title[{\textit Spitzer} IRAC colours of Submillimetre Galaxies]{{\textit Spitzer} IRAC Infrared Colours of Submillimetre-bright Galaxies}
\author[M. S. Yun et al.]{Min S. Yun$^{1}$\thanks{E-mail: myun@astro.umass.edu}, 
Itziar Aretxaga$^{2}$, 
Matthew L. N. Ashby$^{3}$, 
Jason Austermann$^{1}$, 
\newauthor
Giovanni G. Fazio$^{3}$, 
Mauro Giavalisco$^{1}$, 
Jia--Sheng Huang$^{3}$, 
David H. Hughes$^{2}$, 
\newauthor
Sungeun Kim$^{4}$, 
James~D.~Lowenthal$^{5}$, 
Thushara Perera$^{1}$, 
Kim Scott$^{1}$, 
\newauthor
Grant W. Wilson$^{1}$, 
Joshua D. Younger$^{3}$\\
$^{1}$Department of Astronomy, University of Massachusetts, Amherst, MA 01003, USA\\
$^{2}$Instituto Nacional de Astrof\'{i}sica, \'{O}pitca y Electr\'{o}nica (INAOE), Tonantzintla, Peubla, M\'{e}xico\\
$^{3}$Harvard-Smithsonian Center for Astrophysics,
    60 Garden Street, Cambridge, MA 02138, USA\\
$^{4}$Astronomy and Space Sciences Department, Sejong University, 98 Kwangjin-gu, Kunja-dong, Seoul, 143-747, Korea\\
$^{5}$Astronomy Department, Smith College, Northampton, MA 01060, USA}
\begin{document}

\date{Accepted 2008 June 12}
\date{\today}

\pagerange{\pageref{firstpage}--\pageref{lastpage}} \pubyear{2008}

\maketitle

\label{firstpage}

\begin{abstract}
High-redshift submillimetre-bright galaxies identified by blank field surveys at millimetre and submillimetre wavelengths appear in the region of the IRAC colour--colour diagrams previously identified as the domain of luminous active galactic nuclei (AGNs).  Our analysis using a set of empirical and theoretical dusty starburst spectral energy distribution (SED) models shows that power-law continuum sources associated with hot dust heated by young ($\la 100$ Myr old), extreme starbursts at $z>2$ also occupy the same general area as AGNs in the IRAC colour--colour plots.  A detailed comparison of the IRAC colours and SEDs demonstrates that the two populations are distinct from each other, with submillimetre-bright galaxies having a systematically flatter IRAC spectrum ($\ga1$ mag bluer in the observed [4.5]--[8.0] colour).  Only about 20\% of the objects overlap in the colour--colour plots, and this low fraction suggests that submillimetre galaxies powered by a dust-obscured AGN are not common.  The red IR colours of the submillimetre galaxies are distinct from those of the ubiquitous foreground IRAC sources, and we propose a set of IR colour selection criteria for identifying SMG counterparts that can be used even in the absence of radio or \spitzer\ MIPS 24 \micron\ data.  
\end{abstract}


\section{Introduction}

One of the most exciting developments of the past decade has been the resolution of the cosmic far-infrared background into discrete sources, providing a first glimpse of the rapid build-up of massive galaxies in the early universe long predicted by theory.  Deep, wide blank--field surveys at millimetre (mm) and submillimetre (submm) wavelengths \citep{smail97,barger98,hughes98,eales99,scott2002,wang04,wang06,greve04,laurent05,mortier05,bertoldi07,scott08} have shown that ultraluminous infrared galaxies (ULIRGs) at $z\ga 1$ contribute significantly to the observed far-IR background.  Multi--wavelength follow--up studies of these so-called submillimetre galaxies (SMGs) suggest that they are massive, young galaxies seen during the period of rapid stellar mass build-up, with very high specific star formation rates 
at $z>1$ \citep[see review by][]{blain02}.

A major obstacle in understanding the nature of this luminous dusty galaxy population is the limited angular resolution ($>10\arcsec$) of current instrumentation, which prevents unambiguous identification of their counterparts at other wavelengths.  Deep interferometric radio imaging and \spitzer\ 24 \micron\ MIPS imaging are shown to be effective for identifying a subset of SMGs at $z\la3$ \citep{ivison02,chapman05,pope06}, and our understanding of their nature and evolution is based almost entirely on some 100 SMGs identified this way.  Recent 890 \micron\ continuum imaging of the seven brightest AzTEC \citep{wilson08} 1100 \micron\ sources in the COSMOS field \citep{scoville07} using the Smithsonian Submillimeter Array (SMA) by \citet{younger07} has shown that there may be a substantial population of higher redshift ($z>3$) SMGs that are extremely faint or undetected at radio and/or 24 \micron\ MIPS bands.  This discovery accentuates the need for a new method to identify and investigate the higher redshift SMGs in order to quantify their contribution to the cosmic energy budget and star formation history, and to map their evolution over time.    

One common feature among all seven COSMOS AzTEC/SMA sources and the four other  SMGs detected with the SMA by \citet{iono06}, \citet{wang07}, and \citet{younger08} is that they are all detected in the \spitzer\ IRAC bands at $\ga1\mu$Jy level, raising the exciting possibility that deep IRAC imaging may provide a powerful new tool for identifying and obtaining deeper understanding of the SMG phenomenon.  The IRAC bands cover at $z\sim3$ the rest-frame optical to near-IR portion of the spectral energy distribution (SED) and offer direct insight to the properties of their stellar component.  A detailed analysis of the multiwavelength properties of the AzTEC/SMA sources is presented elsewhere (Yun et al., in prep.).  In this paper, we examine the rest-frame optical/near-IR properties of a large sample of well-studied SMGs and the utility of the IRAC colour--colour plots for identifying and investigating the nature of the SMG population in general.

\begin{figure*}
\centering
\includegraphics[width=5.5in]{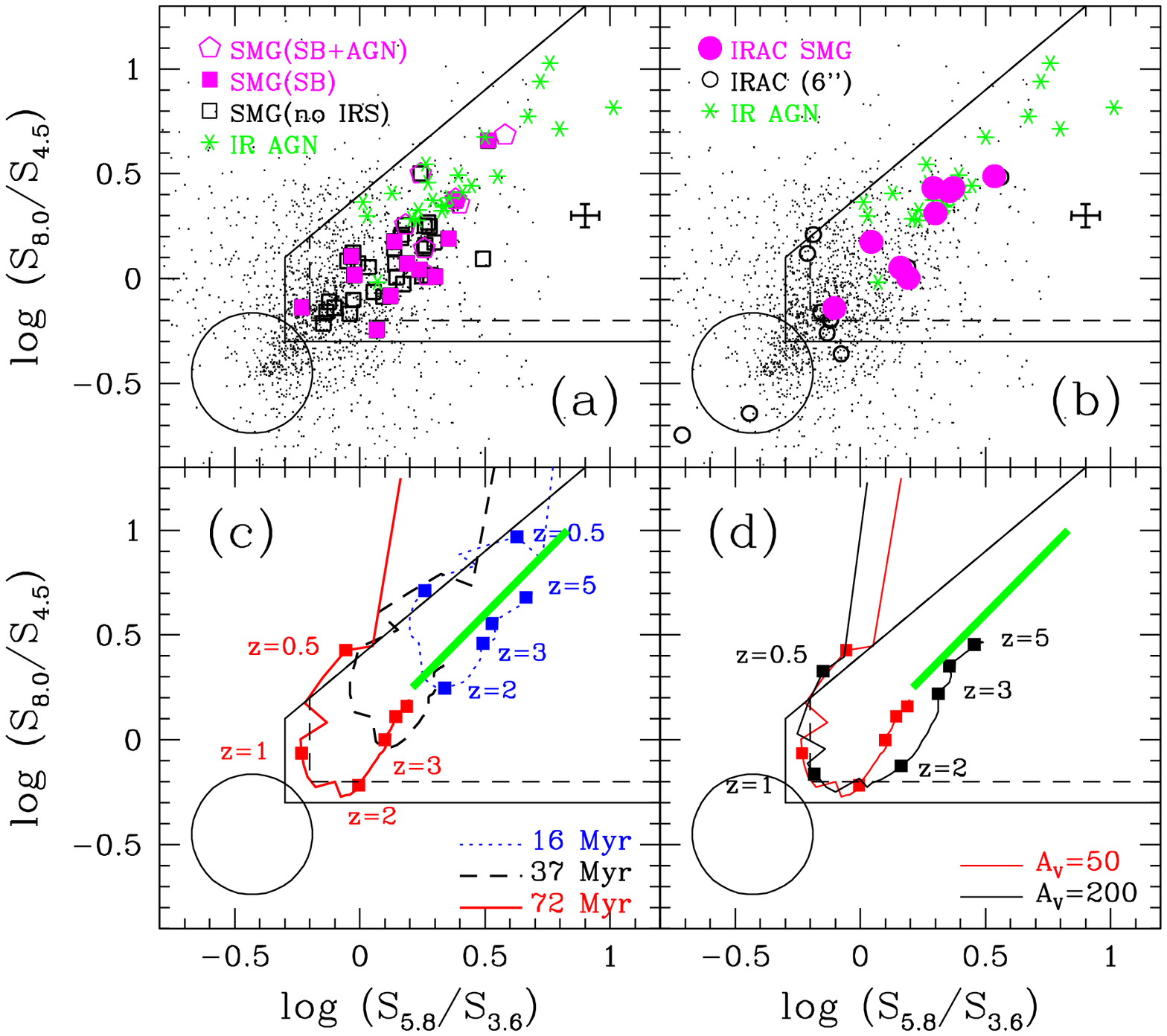}
\caption{{\it Spitzer}/IRAC $S_{5.8}/S_{3.6}$ vs. $S_{8.0}/S_{4.5}$ colour--colour diagram for SMGs and IR QSOs.  
{\bf (a)} Empty squares are SMGs securely identified by radio and CO interferometric imaging.  Filled squares and pentagons are SMGs identified as ``starburst'' and ``starburst+AGN'' by \spitzer\ IRS spectra, respectively \citep{menendez07,valiante07,rigby08,pope08}.  IR QSOs with power-law spectrum identified in the FLS field by Lacy et al. and \citet{martinez08} are shown as stars.  Small dots represent 4000 random field galaxies selected in the COSMOS field \citep{sanders07}, and the large circle centered near ($-$0.4,$-$0.4) represents the centroid of the IRAC sources in the First Look Survey (FLS) field identified by \citet{lacy04}. The long-dashed line outlines the region for AGNs proposed by Lacy et al.  The solid line shows the extended region we propose for the identification of the SMG counterparts (Eq.~1).  An error bar corresponding to a typical 10\% uncertainty in the flux density measurement is shown on the right side.
{\bf (b)} The same {\it Spitzer}/IRAC colour--colour diagram for IRAC sources found within a 6\arcsec\ radius of the nine securely identified mm/submm sources.  Filled circles are the nine SMGs identified by direct submm interferometric measurements obtained using the SMA while the empty circles represent the foreground/interloper IRAC sources.
{\bf (c)} Redshift evolution colour--colour tracks for three different starburst ages based on theoretical starburst SED models by \citet{efstathiou00}.  Filled and empty squares along the redshift evolution colour tracks mark the redshifts of $z=0.5$, 1, 2, 3, 4, and 5.  The thick solid line represents power-law spectrum sources with $\alpha$=0.3-1.0 ($S_\nu \propto \nu^{-\alpha}$).
{\bf (d)} The effects of extinction are demonstrated by the 72 Myr old starburst colour--colour model tracks with total extinction of $A_V=50$ and 200.  
}
\label{fig:colourcolourplot}
\end{figure*} 

\section{\spitzer\ IRAC colours of Submillimetre Galaxies and AGNs}
\label{sec:colourcolour}

\subsection{Red IRAC colours of SMGs}

A colour--colour diagram is a powerful tool for analyzing SEDs sampled coarsely with just a few broad-band measurements.  The four \spitzer\ IRAC bands (3.6 \micron, 4.5 \micron, 5.8 \micron, and 8.0 \micron) probe the portion of the galaxy SED that includes photospheric emission from cool stars (at $z=0-4$) and power-law continuum from hot dust surrounding young stars and/or AGN.  Polycyclic aromatic hydrocarbon (PAH) and other spectral features are also important for the $z\sim0$ galaxies.  The 1.6 \micron\ stellar photospheric PAH feature is prominent in stellar systems older than 10 Myr, and galaxies with substantial cool stellar populations appear blue in an IRAC colour--colour diagram as a result \citep[see][]{simpson99,sawicki02}.  Sources dominated by stellar photospheric emission form a densely concentrated cloud near ($-$0.4, $-$0.4)  in the $S_{5.8}/S_{3.6}$ vs. $S_{8.0}/S_{4.5}$ colour--colour diagram for a sample of field galaxies shown by \citet{lacy04}.  The $z\sim0$ late type galaxies with varying amounts of PAH emission in the 8.0 \micron\ band form a distinct branch of galaxies with a constant $S_{5.8}/S_{3.6}$ ratio emerging from this cloud.  Noting that 54 quasars identified in the Sloan Data Release 1 quasar survey \citep{schneider03} are associated with a second branch with red IR colours, Lacy et al. identified a wedge-shaped area in the colour--colour diagram as the characteristic region populated by obscured and unobscured AGNs.

To explore the importance of AGN activity among SMGs, we examine the IR properties of a sample of 47 well studied SMGs and compare them to AGNs using the AGN diagnostic IRAC colour--colour plot by \citet{lacy04}.  Our SMG sample is constructed from the literature primarily for their secure identification using high angular resolution interferometric imaging at radio wavelengths and deep \spitzer\ MIPS imaging \citep{ivison05,greve05,tacconi06,pope06}.  A small subset of this sample has been observed with the \spitzer\ Infrared Spectrograph \citep[IRS;][]{menendez07,valiante07,rigby08,pope08}, and those classified as ``starburst'' (strong PAH features) and ``starburst+AGN'' (PAH plus power-law continuum) are shown using different symbols in Figure~\ref{fig:colourcolourplot}.  Of these 20 submm-selected galaxies, none shows a pure power-law-dominated IRS spectrum characteristic of an AGN.  We compare the IR colours of our SMG
sample to those of the 19 $z\sim1$ AGNs identified by their power-law continuum in the First Look Survey (FLS) field by \citet{lacy04}, and to the IR colours of the foreground galaxy population, using the 4000 IRAC sources randomly selected from the COSMOS field \citep{sanders07}.   The IRAC colours of these foreground sources are slightly offset from the FLS centroid (indicated in Figure~\ref{fig:colourcolourplot} by a large circle), probably because the COSMOS sources are on average fainter and at higher redshift.    

A surprising result is that more than 90\% of the SMGs are located within the region designated for AGNs by \citet{lacy04}.  It is tempting to interpret this result as an indication that the majority of SMGs host a luminous AGN, as  \citet{greve07} concluded for their MAMBO-selected SMGs in the field surrounding the $z=3.8$ radio galaxy 4C~41.17.   However, the majority (13 out of 20, or 65\%) of SMGs observed with \spitzer\ IRS show spectra dominated by PAH features, characteristic of pure starburst systems.  The remaining seven SMGs show starburst+AGN hybrid spectra, suggesting that the power-law AGN emission is not the dominant contributor to the total IR luminosity of these SMGs either.  Four of the SMGs targeted by \citet{valiante07} were undetected by their \spitzer\ IRS observations.  Deep Chandra X-ray data in the GOODS-North field suggest that only a small fraction of SMGs are detected in the hard X-ray band \citep{pope06,pope08}, further supporting the notion that energetically-dominant, luminous AGNs are not common among the SMGs \citep[also see][]{alexander03}.

\begin{figure*}
\centering
\includegraphics[width=5.5in]{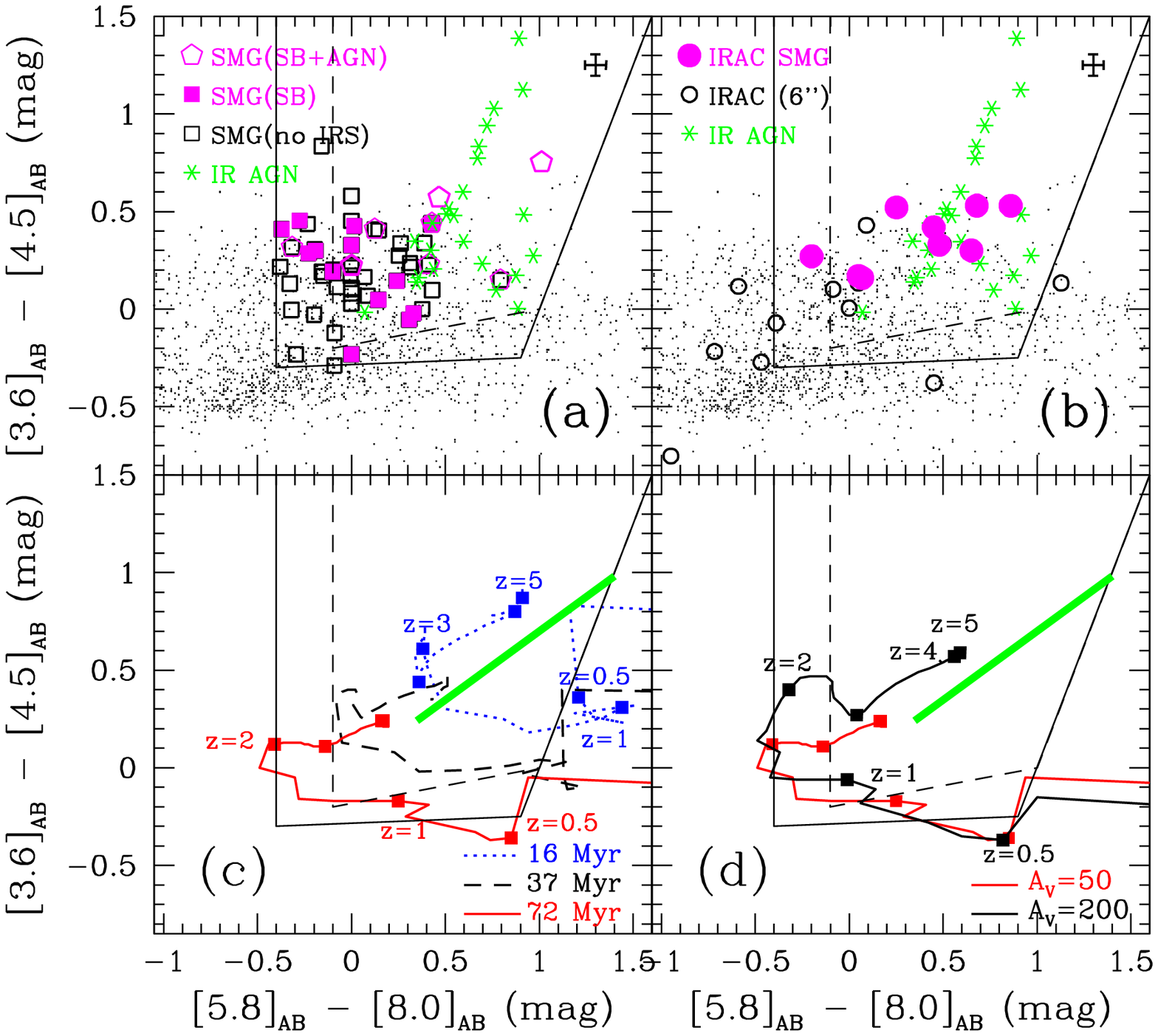}
\caption{{\it Spitzer}/IRAC [5.8]--[8.0] vs. [3.6]--[4.5] colour--colour diagram for submillimetre galaxies and IR QSOs in AB magnitudes.  
{\bf (a)} All symbols and models plotted are identical to those in Figure~\ref{fig:colourcolourplot}. The thin long-dashed line outlined the AGN region proposed by \citet{stern05}.  The solid line shows the extended region we propose for the identification of the SMG counterparts (Eq~2).  
{\bf (a)} The same {\it Spitzer}/IRAC colour--colour diagram for IRAC sources found within a 6\arcsec\ radius of the nine securely identified mm/submm sources (filled circle).  Empty circles are foreground/interloper IRAC sources found within the same search radius as discussed in the text.  The thick-solid line represents power-law spectrum sources with $\alpha$=0.3-1.0 ($S_\nu \propto \nu^{-\alpha}$) in both panels.
{\bf (c) \& (d)} The same theoretical colour--colour evolution tracks by \citet{efstathiou00} shown in Figure~\ref{fig:colourcolourplot} are plotted for comparison.
}
\label{fig:colourcolourplot2}
\end{figure*} 

Another commonly used diagnostic for AGN activity is the {\it Spitzer}/IRAC [5.8]--[8.0] vs. [3.6]--[4.5] colour--colour diagram, first introduced by \citet{stern05}.  In addition to red colours, Stern et al. utilize empirical colour tracks of star-forming galaxies (e.g., M82) to further differentiate starbursts from AGNs.  We reproduce the Stern et al. plot in Figure~\ref{fig:colourcolourplot2} using the same samples as used in Figure~\ref{fig:colourcolourplot}.  About 1/3 of the SMGs (16/47), including 5 out of 13 ``starburst'' IRS spectrum sources, are now found outside the AGN region outlined by Stern et al.  However, a large fraction of the SMGs (31/47), including many with ``starburst'' IRS spectrum, are within the AGN region.

\subsection{Analysis using empirical and theoretical SED models}

A natural explanation for the observed red IR colours of SMGs emerges when empirical and theoretical SED models of dusty starbursts are examined in the context of these IRAC colour--colour plots.  
The extreme IR luminosity of the mm/submm detected sources ($L_{IR}\ge10^{12-13}L_\odot$) indicates that a dust-obscured extreme starburst or a luminous AGN dominates their energy output, and the entire observed SED can be described by a relatively simple model \citep[see][]{yun02}.  Theoretically motivated radiative transfer models for dust-enshrouded starbursts and AGNs \citep{silva98,efstathiou00,siebenmorgen07,chakrabarti08} are successful in reproducing the observed UV-to-radio SEDs, including those of extreme objects such as SMGs and ``hyperluminous'' IR galaxies not commonly found in the local universe \citep[see][]{farrah03,efstathiou03,vega08,groves08}.  
A large number of free parameters (e.g., source geometry, initial mass function, metallicity) and some degeneracy among them limit the utility of these SED models for unique quantitative interpretations of the photometric data.  Nevertheless, they are highly valuable tools for evaluating the effects of model parameters such as starburst age and extinction.  For this work, we adopt the model SEDs computed by \citet{efstathiou00} for their ease of use and because they have already been widely tested \citep[e.g.][]{efstathiou03,farrah03}.   

\begin{figure}
\centering
\includegraphics[width=\hsize]{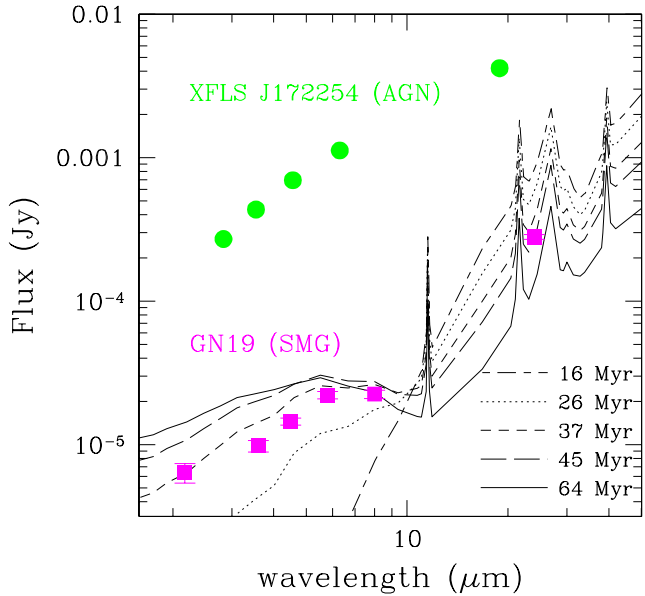}
\caption{A Comparison of the \citet{efstathiou00} model SEDs with the observed SEDs of the $z=1.74$ QSO SSTXFL~J172253.9+582955 \citep{lacy04} and the $z=2.49$ SMG SMM~J123707.7+621411 \citep[GN19;][]{pope06}.  The model SEDs are normalized to roughly match the measured flux densities of GN19 in the IRAC bands. The SED of SSTXFL~J172253.9+582955 is redshifted to $z=2.49$ to match that of GN19 for an easier comparison.  
}
\label{fig:SEDs}
\end{figure}

The Efstathiou et al. SED models are shown in Figure~\ref{fig:SEDs} along with the observed SEDs of the $z=2.49$ SMG J123707.7+621411 \citep{pope06} and the $z=1.74$ QSO SSTXFL J172253.9+582955 \citep{lacy04} in order to illustrate the IRAC colour evolution with starburst age.   The shape of the model SED changes quickly during the first 100 Myr, primarily driven by the rapid evolution of the young stellar population.  As the stellar population ages, the combined effect of decreasing radiation intensity and increasing photospheric emission (``1.6 \micron\ bump'') makes the SEDs flatten (become bluer) in the IRAC bands.  These are generic features of all SED models, only dependent on the details of the input stellar population synthesis model.  Even though the build-up of cool stars and the 1.6 \micron\ photospheric component become evident as early as $\sim$30 Myr after the initial starburst, the overall colour of starburst systems remains {\it red} even at 64 Myr after the burst.  In comparison, the red continuum of the IR AGN increases monotonically across the IRAC bands into the mid-IR (MIPS 24 \micron) band, with a steeper slope than that of the SMG and most of the theoretical starburst SEDs.  The comparison shown in Figure~\ref{fig:SEDs} nicely demonstrates the clear difference in the origin of their near-IR emission and the spectral slope between IR AGNs and SMGs, despite the broad similarity in their red IRAC colours.

Three colour tracks for a single starburst population SED model with different ages are shown in Figures~\ref{fig:colourcolourplot}c \& \ref{fig:colourcolourplot2}c.  They cover the full range of IRAC colours associated with both AGNs and SMGs, with the older starburst tracks showing successively {\it bluer} IRAC colours.  The majority of SMGs appear scattered about the model SED colour tracks, consistent with their \spitzer\ IRS spectra being characteristic of starburst-dominated systems.  For a given model SED, the IRAC colour becomes monotonically redder with increasing redshift at $z\ga1$, and most SMGs have colours consistent with model SEDs redshifted to $z=1-5$.  The colour dependencies on starburst age and redshift are nearly parallel to each other, leading to some degeneracy between the two quantities.  Nevertheless, the observed red IRAC colours of SMGs can arise {\em only if SMGs are at high redshift} ($z\ga1$).  

The effects of extinction (column density along the line of sight) are not as important in determining the observed IRAC colours of SMGs, unlike the case at shorter, optical wavelengths.  The model colour tracks for two different visual extinctions ($A_V=50$ and 200) shown in Figures~\ref{fig:colourcolourplot}d \& \ref{fig:colourcolourplot2}d track each other closely at $z\la2$, despite the large opacity difference between the two models.   Optical depth is greatly reduced at these long wavelengths \citep[$A_{\lambda}/A_V=20\sim40$ for the IRAC bands; see][]{indebetouw05,roman07}, and the weak dependence on extinction $A_V$ can be naturally understood.  As redshift increases, the IRAC bands begin to probe the near-IR to optical bands, and an increasing dependence on extinction is expected.  Indeed the IRAC colours of the two model SEDs diverge at $z>2$, with the higher extinction $A_V=200$ model predicting redder IRAC colours as expected (see Fig.~\ref{fig:colourcolourplot}d), and some degeneracy between redshift and extinction is also identified.  Overall, extinction still plays a minor role compared with starburst age, and we identify starburst age and redshift as the dominant physical parameters that affect the observed IRAC colours.  Accounting for SMGs with the reddest observed colours (e.g., log($S_{5.8}/S_{3.6}$)$>$0.3, log($S_{8.0}/S_{4.5}$)$>$0.3) requires young ($\la30$ Myr old) starbursts at high redshifts ($z\ga3$) or a power-law AGN dominating the rest-frame near-IR SED.

An AGN-like warm IR colour is a generic feature for {\it all} stellar systems whose near-IR luminosity is dominated by a dust-obscured young stellar population.  In analyzing the ``mid-IR excess'' of galaxies selected using the Infrared Astronomy Satellite (IRAS) data, \citet{yun01} noted that the youngest dusty starbursts can display warm mid-IR colours, exceeding the classic Seyfert division at log $(S_{25\mu}/S_{60\mu}) \ge 0.18$ \citep{degrijp85}.  While exploring the presence of power-law AGN candidates in the Chandra Deep Field North region, \citet{donley07} also noted the incursions by their empirical ULIRG colour tracks (derived from the observed SEDs of Arp~220, IRAS~17208$-$0014, and Mrk~273) into the AGN region in their IRAC colour--colour diagram.  The incursion of M82-like objects into the AGN boundary has lead \citet{stern05} to refine the  selection boundary, and \citet{barmby06} question the completeness and reliability of AGN identification using IRAC colours for a similar reason.  Using theoretical and empirical starburst SED models, we demonstrate that young, dusty starbursts exhibit red IRAC colours, and red IRAC colour is {\it not} unique to power-law AGNs.  The popular AGN identification methods using red IRAC colours, such as by \citet{lacy04} and \citet{stern05}, should be used with caution and a clear understanding of this important caveat.

\subsection{Systematic colour difference between SMGs and AGNs}

Both SMGs and AGNs appear within the broadly defined red IRAC colour regions by Lacy et al. and Stern et al., but SMGs and AGNs are distinguished by clear systematic difference in their mean IRAC colours.  The SMGs as a group are systematically bluer than the AGNs, with only about 20\% of SMGs appearing mixed in among the power-law AGNs in both diagnostic colour--colour plots.  All AGNs show log($S_{8.0}/S_{4.5}$) $\ge0.2$ in Figure~\ref{fig:colourcolourplot} while the overwhelming majority ($\ga80\%$) of the SMGs have a smaller flux ratio.  Similarly, all AGNs have [5.8]--[8.0] colours redder than +0.2 in Figure~\ref{fig:colourcolourplot2}.  The most significant colour difference is between the 4.5 \micron\ and 8.0 \micron\ bands with an average difference of $\langle[4.5]-[8.0]\rangle\ga 1$ mag.  The [3.6]--[4.5] colour difference is the smallest, with $\langle[3.6]-[4.5]\rangle \sim 0.25$ mag.   

\begin{figure}
\centering
\includegraphics[width=\hsize]{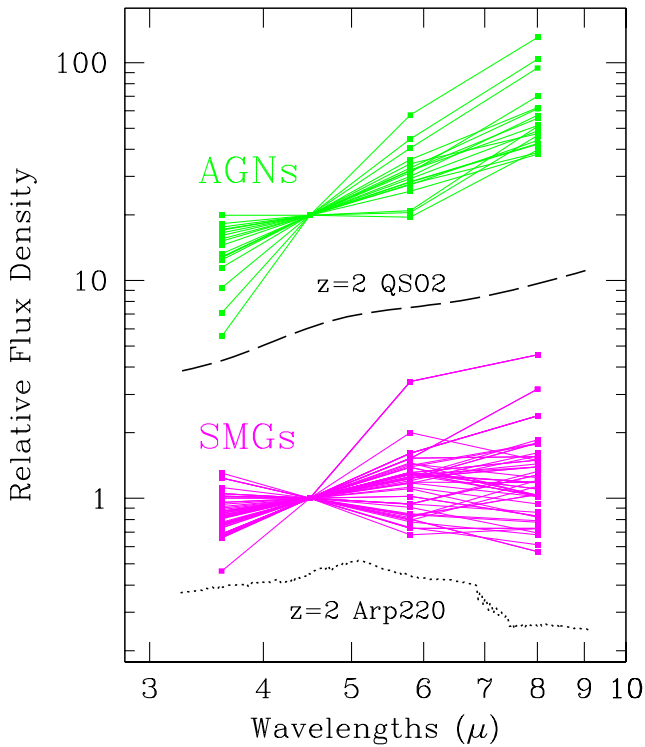}
\caption{The IRAC band continuum spectral energy distributions for the AGNs and SMGs shown in Figures~\ref{fig:colourcolourplot} \& \ref{fig:colourcolourplot2}.  All SEDs are normalized at 4.5 \micron\ band, and the AGNs and SMGs are displaced vertically by an arbitrary amount for ease of comparison.  The $z=2$ SEDs of type-2 QSO and Arp~220 (M82 SED is nearly identical) compiled by \citet{polletta07} are shown for comparison, with an arbitrary offset in the flux scaling.}
\label{fig:IRACSEDs}
\end{figure}

The comparison of the full IRAC band SEDs shown in Figure~\ref{fig:IRACSEDs} offers a visually compelling demonstration of a steeper spectral slope for the AGNs.  The average IRAC SED of the AGN sample is in excellent agreement with the empirical type-2 QSO SED template by \citet{polletta07}. The template SED for a type-1 QSO is nearly identical (not shown) to type-2 QSO at these wavelengths, and the steep spectral slope is characteristic of all obscured and unobscured AGNs.  In comparison, the average IRAC SED of the SMG sample is significantly flatter, with only a few outliers with an AGN-like steeper spectral slope.  The ULIRG Arp~220 SED redshift to $z=2$ offers a far better match to the SMGs, lending further support for the dust obscured starburst interpretation.  Furthermore, the flatter spectral slope (bluer IRAC colours) suggests that SMGs in general do not host a dust-obscured, energetically dominant AGN in most cases.  We note that rigorous diagnostics of an AGN include the detection of high-ionization, high-excitation emission lines and copious amounts of hard X-ray emission, and an IR power-law spectrum is only indirect evidence for the presence of an AGN, and does not directly measure the AGN accretion power.

The distinct difference in the observed SEDs between AGNs and SMGs also offers an important constraint on the possible SMG-QSO evolutionary scenario.  The apparent correlation observed between the black hole mass ($M_{BH}$) and the host spheroid mass \citep[velocity dispersion $\sigma$; see][]{magorrian98,ferrarese00,gebhardt00} has raised considerable interests in the process that leads to the build-up of the stellar mass in galaxies and the growth of the central super-massive black hole.  The evolutionary scenario that a massive nuclear starburst associated with an ultraluminous infrared galaxy leads to a QSO phase \citep{sanders88,norman88} offers an attractive explanation for the $M_{BH}-\sigma$ relation if the SMG phase corresponds to the period of rapid stellar mass build-up.  However, only the minor overlap between the AGNs and the SMGs we find here suggests that the duration of the transition period should be much shorter than the SMG or the IR AGN phase in such an evolutionary scenario.

\section{Identification of Optical/IR Counterparts to Submillimetre Sources}
\label{sec:ID}

As discussed above, every SMG identified with high angular resolution interferometric imaging using the SMA has a clear IRAC counterpart, including those undetected in the radio and in the MIPS 24 \micron\ bands.  \spitzer\ IRAC data may therefore offer a powerful new method for identification of SMGs that is otherwise extremely difficult because of their faintness at optical and near-IR wavelengths.  The uncertainty in the mm-wave positions of SMGs is typically 5-10\arcsec, and the high IRAC source density \citep[$\sim60$ arcmin$^{-2}$ at the 1.4 $\mu$Jy level in the 3.6 \micron\ band;][]{fazio04} is too high to enable unique identification of SMG counterparts in general \citep[see discussions by][]{pope06}.  The new knowledge obtained from Figures~\ref{fig:colourcolourplot}~\&~\ref{fig:colourcolourplot2} that SMGs as a population have red IR colours, similar to AGNs and distinct from the foreground field population, offers the exciting possibility of identifying or significantly narrowing down the counterpart candidates using the IRAC data alone.

The distribution of SMG IRAC colours and the starburst model SED colour tracks in Figures~\ref{fig:colourcolourplot}~\&~\ref{fig:colourcolourplot2} extends slightly beyond the AGN selection boundaries proposed by \citet{lacy04} and \citet{stern05}.   We propose a new set of SMG counterpart candidate selection criteria for the $S_{5.8}/S_{3.6}$ vs. $S_{8.0}/S_{4.5}$ colour--colour diagram, expanded to include all SMGs and the model SED tracks as shown with long-dashed lines in Figure~\ref{fig:colourcolourplot}.  The new proposed criteria are:
\begin{eqnarray}
   log(S_{8.0}/S_{4.5}) > -0.3 &\wedge &
   log(S_{5.8}/S_{3.6}) > -0.3 \nonumber \\
                              & \wedge &
   log(S_{8.0}/S_{4.5}) < log(S_{5.8}/S_{3.6}) + 0.4
\end{eqnarray}
where $\wedge$ is the logical AND operator.  Equivalent criteria for the Stern et al. colour--colour diagram (Fig.~\ref{fig:colourcolourplot2}) in AB magnitudes are:
\begin{eqnarray}
([5.8] - [8.0]) &>&  -0.4 \nonumber \\ 
~~~~~~& \wedge & ([3.6] - [4.5]) > 0.036 \times ([5.8] - [8.0]) - 0.318 \nonumber \\
~~~~~~& \wedge & ([3.6] - [4.5]) > 2.5 \times ([5.8] - [8.0]) - 2.5
\end{eqnarray}
The relative merits of these selection criteria are discussed further below.

To evaluate the utility of these new IRAC colour selection criteria for the SMG counterpart candidate selection, we examine the colours of the IRAC sources in the fields surrounding the nine SMGs securely identified through submm interferometric imaging \citep{iono06,wang07,younger07,younger08}.  The 20\arcsec\ $\times$ 20\arcsec\ IRAC 3.6 \micron\ images (not shown) centered on the original submillimetre source coordinates contain up to a dozen IRAC sources each, demonstrating the difficulty of using the IRAC source catalog alone for the SMG counterpart identification.  There are a total of 20 IRAC sources located within a 6\arcsec\ radius (typical positional accuracy) of the nine nominal SMG positions; their colours are shown in Figures~\ref{fig:colourcolourplot}b~\&~\ref{fig:colourcolourplot2}b.  Colours of the nine SMGs, shown as filled circles, fall within our proposed SMG colour boundaries in both plots.  These SMGs were not used in defining the IRAC colour selection criteria and they thus represent an independent and successful test of our method.  The remaining IRAC sources have colours consistent with the background sources, and about 1/2 of them fall within the new SMG candidate identification boundaries.  The new colour selection technique does not completely resolve the confusion problem.   However, the situation is now greatly improved, as the candidate counterparts are reduced to a single unique candidate or two potential candidates, making expensive follow-up observations more efficient.  Other identification methods such as deep radio or MIPS 24 \micron\ imaging do not greatly improve the situation in these test cases since many of these SMGs are undetected at other wavelengths. Three of the securely identified sources have IRAC colours typical of $z\sim2$ starburst systems while the remaining six cluster near the power-law AGNs (see Fig.~\ref{fig:colourcolourplot}b \&~\ref{fig:colourcolourplot2}b).  In addition to being undetected in the radio, these six SMGs with AGN-like colour are undetected in the \spitzer\ MIPS 24 \micron\ bands, ruling out the presence of a power-law AGN component that extends into the mid-IR bands, as seen for the QSO spectrum shown in Figure~\ref{fig:SEDs}. Therefore these SMGs with very red IRAC colours may be the youngest, highly-obscured starburst systems at very high redshifts \citep[$z\ga3$; also see the discussions in][]{younger07}.

The IRAC colour selection method described here is qualitatively similar to the IRAC 8.0~\micron\ band selection method used to identify the SCUBA galaxies in the CUDSS 14 hour field by \citet{ashby06}.  Noting the red colour of the SMGs and reduced foreground confusion compared with the shorter wavelength IRAC bands, Ashby et al. argued that 8.0 \micron\ selection offers a better means for identifying SMG counterparts than near-IR or optical selection.  Their 8.0 \micron\ selection resulted in a different counterpart from the optical or $K$-band selection methods in ten out of 17 cases, casting some doubt on the earlier identifications of SMGs.  The main difference of the new IRAC colour selection method outlined here is that we fully quantify the red colour of the SMG population using Eqs.~1 \& 2 for easy implementation and calibrate them using a large, well-studied sample in conjunction with empirical and theoretical dusty starburst SEDs.  

\section{Discussion and Concluding Remarks}
\label{sec:summary}

High-redshift submillimetre-bright galaxies identified by blank field surveys at millimetre and submillimetre wavelengths appear in the region of the IRAC colour--colour diagrams previously identified as the domain of luminous AGNs by \citet{lacy04} and \citet{stern05}.  Rather than interpreting this as a sign that the majority of the SMGs are powered by a luminous AGN, we have shown using empirical and theoretically motivated dusty starburst SED models that their IR colours can be interpreted as those of power-law continuum associated with hot dust heated by young ($\la 100$ Myr old), extreme {\it starbursts} at $z>2$.  These SMGs fall along the branch of galaxies extending from the blue photospheric peak toward the red power-law region in the IRAC colour--colour plot by \citet{lacy04}, and our analysis suggests that this branch is a heterogeneous ensemble of power-law AGNs and dust obscured starbursts whose continuum near-IR luminosity exceeds that of the photospheric emission from cool stars.  In fact, our analysis demonstrates that the popular red IRAC colour selection methods for AGN identification, such as the criteria by Lacy et al. and Stern et al., should be used with caution because of the significant starburst contribution expected.  While there is some overlap between SMGs and AGNs in these IRAC colour--colour plots, SMGs are systematically bluer ($\langle[4.5]-[8.0]\rangle\ga 1$ mag), consistent with 30 to 70 Myr old starbursts observed at redshifts between $z\sim1$ and $z\sim5$.  

Our examination of the model dusty starburst SEDs of \citet{efstathiou00} shows that the main physical parameters that determine the observed IRAC colours are starburst age and redshift, while extinction (column density) plays a less important role.  For the models examined in Figures~\ref{fig:colourcolourplot} and \ref{fig:colourcolourplot2}, starburst age and redshift are nearly degenerate.  The use of IRAC colours as a redshift indicator was first proposed by \citet{simpson99} and \citet{sawicki02}, and various empirical photometric redshift relations have been proposed recently with estimated accuracies of $\delta z/(1+z)=0.1-0.2$ \citep[see][]{pope06,wilson08b}.   This photometric redshift method may not work well for galaxies with red power-law continuum and/or weak 1.6 \micron\ photospheric feature that are more common at $z\ga2$.  The degeneracy between the starburst age and redshift found with these theoretical model SEDs suggests that a careful calibration using a large sample of SMGs is needed in order to quantify fully the systematic uncertainty of this method.

A detailed comparison of the IRAC colour--colour plots and SEDs shows that AGNs and SMGs are distinct from each other due to intrinsic differences in their energy source and dust distribution.  SMGs as a group have a flatter SED (bluer by $\langle[4.5]-[8.0]\rangle\ga 1$ mag) in comparison with AGNs.  Only 20\% of the objects overlap in the colour--colour plots shown in Figures~\ref{fig:colourcolourplot}~\&~\ref{fig:colourcolourplot2}, and this suggests that SMGs powered by an AGN are not common.  In the context of the ULIRG-QSO evolutionary scenario \citep{sanders88,norman88}, the little overlap between the AGNs and the SMGs population may indicate that the transition period is much shorter than the duration of the SMG or the IR AGN phase.

The red IR colours of the SMGs are distinct from the colours of the majority of the foreground IRAC sources, which are both early and late type galaxies at $z\la2$.  We show that colour selection criteria similar to those of AGNs proposed by Lacy et al. and Stern et al. can be used to pinpoint the IRAC counterpart to an SMG uniquely or to pare down the candidates to just a few, improving the efficiency of expensive follow-up observations at other wavelengths.  The lack of high-quality multi-wavelength data for a large sample of SMGs is currently the primary limiting factor that prevents a better understanding of the SMG phenomenon.  The IRAC colour selection method discussed here appears to work even in cases when the radio and MIPS 24 \micron\ counterparts are too faint to be detected at the present sensitivity, and an unbiased investigation of the entire SMG population may become possible using this new identification method. 

There are some important advantages and disadvantages to using the ($S_{5.8}/S_{3.6}$ vs. $S_{8.0}/S_{4.5}$) colour--colour plot (Figure~\ref{fig:colourcolourplot}) versus the ([5.8]--[8.0] vs. [3.6]--[4.5]) colour--colour plot (Figure~\ref{fig:colourcolourplot2}) for SMG candidate identification.  The former uses IR colours with a longer baseline in wavelength, making the colour measurements more robust and the model colour tracks better behaved, as shown in Figure~\ref{fig:colourcolourplot}.  On the other hand, the instrumental sensitivity of the IRAC 3.6 \micron\ and 4.5 \micron\ bands are a factor of a few better than the 5.8 \micron\ and 8.0 \micron\ bands, and the 5.8 \micron\ and 8.0 \micron\ channels will be inoperable during the upcoming warm \spitzer\ operation.  The model colour tracks shown in Figure~\ref{fig:colourcolourplot2} are more irregular because of shorter wavelength baselines in these colours, but the [3.6]--[4.5] colour is the best determined quantity among all IRAC colour combinations.  

A far simpler and potentially more robust alternative SMG candidate selection criterion is [3.6]--[4.5] $>-0.2$.  This single colour selection method does only a slightly poorer job of rejecting foreground sources than the full colour section criteria described by Eq.~2.  Given the superior sensitivity of the IRAC 3.6 \micron\ and 4.5 \micron\ bands and their availability during the warm \spitzer\ mission, this simpler colour selection criterion may be the more effective method for identifying SMG counterpart candidates in the longer term.  Alternatively, colours of IRAC to optical or near-IR bands or to MIPS 24 \micron\ bands have also been proposed previously for identifying AGN activity \citep[see][]{huang04,ashby06,webb06}, and similar analysis may also prove fruitful for SMG candidate identification.

\section*{Acknowledgments}

We thank M. Poletta for providing us with the library of empirical galaxy SED templates used in Figure~\ref{fig:IRACSEDs}.  This work is based on observations made with the {\it Spitzer} Space Telescope, which is operated by the Jet Propulsion Laboratory, California Institute of Technology, under NASA contact 1407.  This work is also partially funded by NSF Grant AST 05-40852 to the Five College Radio Astronomy Observatory.

\bsp

\label{lastpage}

\end{document}